**Quality factor due to roughness scattering of shear horizontal surface acoustic waves in nanoresonators**

G. Palasantzas [a]

Zernike Institute of Advanced Materials, University of Groningen, Nijneborgh 4, 9747 AG Groningen, The Netherlands

**Abstract**

In this work we study the quality factor associated with dissipation due to scattering of shear horizontal surface acoustic waves by random self-affine roughness. It is shown the quality factor is strongly influenced by both the surface roughness exponent $H$, and the roughness amplitude $w$ to lateral correlation length $\xi$ ratio. Indeed, quality factors for roughness exponents $H \geq 0.5$ are comparable to quality factors due to intrinsic dissipation mechanisms (e.g., thermoelastic losses and phonon-phonon scattering) especially for wave vectors $<1/\xi$. Our results indicate that this dissipation mechanism should be carefully considered in the design micro/nanoelectromechanical systems.

*Pacs numbers:* 68.55.Jk, 85.85.+j, 68.55.-a, 73.50.Td

___________________________________

[a] Corresponding author: G.Palasantzas@rug.nl

## I. Introduction

A central theme in the development of micro/nano-electromechanical systems (MEMS/NEMS) is to achieve high-quality factors in micro/nanometer sized mechanical oscillators, which are important for both fundamental (where either the mechanical system or its interaction with the environment must be described quantum mechanically) and applied research [1-12]. It has been well established that this is directly related to the sensitivity of many sensor applications, where it is of primary importance to understand various dissipation mechanisms of mechanical energy and to minimize them. In general, the quality factor $Q$ of a mechanical oscillator is determined by the various multiple loss mechanisms (assuming independent) as

$$Q^{-1} = \sum_i Q_i^{-1}. \tag{1}$$

The index $i$ denotes loss due to environment (gas molecules impinge the resonator surface), losses due to bulk defects and impurities, losses due to surface effects, losses due to the thermoelastic effects (thermal currents generated by vibratory volume changes in elastic media with nonzero thermal-expansion coefficient), and losses due to phonon-phonon scattering etc. [1-5, 11, 12].

In a variety of studies it has been shown that surface roughness decreases the quality factor under operation at low pressure environment. Studies for SiC/Si NEMS indicted that devices operational in the UHF/microwave regime had a low surface roughness (~ *2 nm*), while devices with rougher surfaces (~ *7 nm*) were operational only up to the VHF range [12]. For Si nanowires it was shown that the quality factor was drastically decreased by increasing the surface to volume ratio [8, 9]. Recently, within the molecular regime (molecule mean free path larger than the lateral nanoresonator dimensions), it was shown that surface roughness can decrease the quality factor and dynamic range, and increase the limit to mass sensitivity [13, 14].

Nevertheless, up to now the quality factor associated with scattering of surface sound waves by self-affine random roughness, which is observed in a variety of surface treatment and growth techniques [14], remains unexplored. Moreover, it is necessary to gauge properly the contribution of sound wave scattering on the quality factor in comparison to dissipation by fundamental energy losses. For our purpose we shall consider initially the case of shear horizontal (SH) surface waves, which are also used in acoustic wave sensors. Notably SH-based sensors have proven to be powerful devices for detection and quantification of different physical parameters as for example mass loading since they offer high enough sensitivity [15].

**II. Brief theory of scattering and quality factor**

Furthermore, we consider the propagation of SH waves along the X-axis with displacement field given by $u = (0, u(X, Z | \omega) \exp(-i\omega t), 0)$ in the region $Z > h(X)$ with h(X) the one-dimensional roughness profile (with X the in-plane position along the rough profile, see Fig. 1) [16]. Effectively we consider here the roughness along the direction of sound propagation only as in ref. 16, while a more general treatment has to take into account the full two dimensional case – a rather complex problem (see in [16] refs. therein). The field u(X,Z|ω) satisfies the wave equation

$$-\omega^2 u = c_t^2 (\partial_X^2 + \partial_Z^2) u \qquad (2)$$

where $\vartheta_X \equiv \partial/\partial X$ are the partial derivatives. Equation (2) obeys the boundary conditions of a stress free surface

$$-(\vartheta_X h)\vartheta_X u + \vartheta_Z u |_{Z=h(X)} = 0 \text{ and } u(X,Z|\omega)|_{Z\to\infty}=0. \tag{3}$$

$c_t$ is the velocity of sound into the solid [16]. By substituting a solution of the form $u(X,Z|\omega) = \int A(k,\omega)\exp[-ikX - a_t(k,\omega)Z](dk/2\pi)$ in the boundary conditions of Eq. (3), then the dispersion relation $\omega=\omega(k)$ is obtained. Note that $a_t(k,\omega) = \sqrt{k^2 - \omega^2/c_t^2}$ if $k^2 - \omega^2/c_t^2 > 0$ and $a_t(k,\omega) = -i\sqrt{\omega^2/c_t^2 - k^2}$ if $k^2 - \omega^2/c_t^2 < 0$ [16]. In the limit of weak roughness (or $|\nabla h|<1$), the dispersion relation reads of the form [16]

$$\omega = c_t k[1 - (2k^6/\pi^2)(\Delta_1^2 - \Delta_2^2)] + ic_t k[(4k^6/\pi^2)(\Delta_1 \Delta_2)], \tag{4}$$

with

$$\Delta_1 = \int_0^{+\infty} s^{3/2} C(2ks)/\sqrt{s+1}\,ds + \int_1^{+\infty} s^{3/2} C(2ks)/\sqrt{s-1}\,ds$$
$$\Delta_2 = \int_0^1 s^{3/2} C(2ks)/\sqrt{s-1}\,ds \tag{5}$$

The surface sound wave is scattered only into bulk elastic waves leading to attenuation or non zero imaginary part in Eq. (2) [16]. Indeed, the surface sound wave exists only due to presence of roughness which leads to wave slowing or equivalently real part

*Re[$a_t(k,\omega)$]>0* [16]. Since the imaginary part *Im[$\omega$]* indicates attenuation, we define the associated quality factor by [11]

$$Q_s = \text{Re}[\omega]/2\,\text{Im}[\omega] = [1-(2k^6/\pi^2)(\Delta_1^2-\Delta_2^2)]/[(8k^6/\pi^2)(\Delta_1\Delta_2)]. \tag{6}$$

**III. Results-Discussion**

Calculations of the quality factor $Q_s$ from Eqs. (5) and (6) requires knowledge of the roughness spectrum *C(k)*, which is the Fourier transform of the height correlation function $<h(X)h(0)> = \int C(k)\exp(-ikX)[dk/2\pi]$ [14, 17]. A wide variety of surface treatments lead to surfaces that possess the so-called self-affine roughness [14]. In this case, the roughness spectrum for one-dimensional height profiles (considered along the sound propagation) scales as [14]

$$C(k) \propto \begin{cases} k^{-1-2H} & \text{if } k\xi \gg 1 \\ \\ \text{const} & \text{if } k\xi \ll 1 \end{cases}. \tag{7}$$

The self-affine scaling in Eq. (7) can be described by the simple analytic model [17]

$$C(k) = (Aw^2\xi)/(1+k^2\xi^2)^{(1+2H)/2} \tag{8}$$

with $A = \pi/2\int_0^{Q_c} \xi(1+k^2\xi^2)^{-(1+2H)/2}dk$. The latter is derived under the normalization constraint that the real space height-height correlation function $<h(X)h(0)>$ should give at zero separation $w^2$ or equivalently $\int_{-Q_c \leq k \leq Q_c} C(k)\exp(-ikX)[dk/2\pi]|_{X=0} = w^2$. Small values of the roughness exponent $H$ (~0) characterize jagged or irregular surfaces, while large values of $H$ (~1) surfaces with smooth hills-valleys (see Fig. 1) [14].

If we substitute Eqs. (2) and (3) into Eq.(6) we obtain the more intuitive form for the dispersion relation and the associated quality factor $Q_s$

$$\omega = c_t k \{1 + (w^4/\xi^4)\omega_1(x) - i(w^4/\xi^4)\omega_2(x)\}$$
$$Q_s = [(\xi^4/w^4) + \omega_1(x)]/\omega_2(x) \qquad (9)$$

with $\omega_1(x) = -(2x^6/\pi^2)(\overline{\Delta}_1^2 - \overline{\Delta}_2^2)]$ and $\omega_2(x) = (4x^6/\pi^2)(\overline{\Delta}_1\overline{\Delta}_2)$. Moreover, we have for the functions $\overline{\Delta}_{1,2}$ the expressions:

$$\overline{\Delta}_1 = 2A\int_0^{+\infty}(1+4x^2\cosh^4\theta)^{-(1+2H)/2}\cosh^4\theta d\theta +$$
$$\qquad\qquad 2A\int_0^{+\infty}(1+4x^2\sinh^4\theta)^{-(1+2H)/2}\sinh^4\theta d\theta \qquad (10)$$
$$\overline{\Delta}_2 = 2A\int_0^{\pi/2}(1+4x^2\sin^4\theta)^{-(1+2H)/2}\sin^4\theta d\theta$$

with $x=k\xi$. From Eq.(9) it can be observed that the shear horizontal waves display the phenomenon of wave slowing or $\omega_1<0$ (under also the constraint $(w^4/\xi^4)+\omega_1(x)>0$ since we must have $Q_s>0$ from Eq. (6)), which is what it binds the surface acoustic waves to the surface [16]. In any case, within the approximation of the theory for weak roughness, the existence of shear horizontal surface sound waves is solely due to their scattering by roughness into bulk elastic waves [16]. To quantify further the limit of weak roughness we consider as a more quantitative measure the rms local slope $\rho_{rms} = <|\nabla h|^2>^{1/2} = [\int_0^{Q_c} C(k)k^2 dk]^{1/2} < 1$, which can be computed in terms of the analytic model for *C(k)* from Eq. (8).

In the following, our calculations were performed for roughness amplitudes *w*, which are observed in real experimental NEMS surfaces (typical in the range *w~1-10 nm*) [8, 9, 12]. Figure 2 shows calculations of both *-$\omega_1$, $\omega_2$* vs. *x* for two relatively different roughness exponents *H*, which they can be measured experimentally (typical accuracy *±0.05*) [14]. In comparison with the case of Gaussian roughness [16], for power law roughness a surface sound wave exists for values of *x>1* since *-$\omega_1$>0* in Fig.2. Moreover, we should also have $(\xi^4/w^4) + \omega_1(x) > 0$: for *w/$\xi$≤0.1*, which is typical for random rough surfaces, this condition is satisfied for *0.5<H≤1* for a significant regime of values *x>1*. In any case, with decreasing exponent *H* or increasing surface roughening (at short length scales or *<$\xi$*) the range of existence of surface sound wave increases for values of *x (=k$\xi$)>1*.

Furthermore, Fig. 3 show calculations of the quality factor $Q_s(x)$ vs. *x* for various roughness exponents *H*. $Q_s(x)$ decreases drastically by many orders of magnitude as x increases in the range of values *x<1*. However, for *H=0.5* a drastically different behavior develops for *x>1* where we obtain *Re[$\omega(k)$]<0* leading to a sharp change of the quality factor. If we substitute into $\bar{\Delta}_1$ the general expression $C(k) \propto k^{-(1+2H)}$ (*k$\xi$>>1*) for one-dimensional self-affine spectrum, from Eq. (4), which is also satisfied by analytic model under consideration, then for large values of *θ (>>1)* we obtain $\bar{\Delta}_1 \propto \int^{+\infty} e^{(2-4H)\theta} d\theta$. The latter indicates that $\Delta_1$ becomes divergent for exponents *H≤0.5* upon integration. This occurs even if we consider *w<<$\xi$* to reassure that we preserve during the calculations the condition of weak surface roughness (or equivalently small local surface slopes $|\nabla h|<1$), which underlies the validity of the present formalism).

For Gaussian roughness $C(k) \propto \exp(-vk^2)$ which was considered in the past [16], this divergence does not arise due to its fast decay. Notably, the Gaussian correlation corresponds to roughness exponent $H=1$ if it is considered in terms of the stretched exponential correlation function $<h(X)h(0)> \sim exp[-(X/\xi)^{2H}]$, which has been used widely to describe surface roughness [16, 18]. At any rate, the divergence of $\bar{\Delta}_1$ suggests that one-dimensional self-affine rough profiles can not sustain surface sound waves for roughness exponents $H<0.5$ (antipersistent roughness) [14, 19]. Indeed, if $\bar{\Delta}_1 \to \infty$, then we obtain $\omega_{1,2}(x) \to \infty$ and $\omega_1/\omega_2 \to \infty$ (since $\omega_1 \propto \bar{\Delta}_1^2$), yielding $Q_s \to \infty$ indicating that no dissipation takes place [19].

The quality factor decreases with increasing surface roughening or equivalently decreasing roughness exponent $H$ and/or increasing ratio $w/\xi$ since it leads to stronger scattering of surface sound waves into bulk elastic waves. If we compare Fig. 3 and 4 it is evident that for a change of the roughness exponent in the range $0.5<H\leq 1$ the quality factor $Q_s$ changes comparably in magnitude (or even more) with changes due to different correlation lengths $\xi$ and roughness amplitudes $w$ as shown in Fig. 4. Therefore, it becomes evident that the influence of the particular roughness exponent $H$ is of crucial importance as that of the long wavelength ratio $w/\xi$.

Finally, it is important to compare at least qualitatively (since we consider dissipation only along due to scattering along the sound propagation) $Q_s$ to quality factors from mechanisms setting their upper bounds in MEMS/NEMS. Indeed, a variety of intrinsic dissipation mechanism contributes to the quality factor. The intrinsic dissipation sources include energy dissipation in a perfect crystal lattice (fundamental loss mechanisms), and dissipation in a real imperfect crystal due to bulk and surface defects. The intrinsic mechanisms for a perfect crystal impose the upper bounds to attainable $Q$. These processes include thermoelastic damping arising from anharmonic coupling between mechanical modes and the phonon reservoir [10, 11], and losses due to electron–phonon and phonon–phonon interactions [10]. At any rate, it has been shown that the thermoelastic losses yield quality factors $Q_{th} \sim 10^3 - 10^{10}$ for Si and GaAs beams at various temperatures, while the phonon-phonon interaction was shown to give in general also very large quality factors $Q_{ph-ph} > 10^8$ (depending on the mode of vibration) [10, 11]. If we compare the sound wave quality factor $Q_s$ to that of thermoelastic $Q_{th}$ and phonon-phonon $Q_{ph-ph}$ dissipation, then it becomes apparent that there is a significant range where these quality factors become comparable depending also on the particular system sizes and surface roughness parameters. Moreover, if we consider scattering from a two dimensional surface we expect higher dissipation and thus lower quality factor $Q_s$ indicating that dissipation along the one-dimensional rough profile along the sound propagation sets for this mechanism a lower limit for $Q_s$ [19].

It is also important to point out that the theoretical calculations presented are based on a self-affine rough surface without boundary confinement (or effectively for sound wavelengths much smaller than the system size). Therefore, our analysis can predict significant influence on the quality factor due to surface sound waves when the resonator size along the propagation direction is larger than sound wavelength. In practice since the wavelength of surface sound waves is typically in the range *1 - 100 μm* [20], the resonator size must be above 1 μm (along the propagation direction and frequencies less than *GHz*). However, for operating frequencies above *1 GHz* the resonator size can be further reduced.

**IV. Conclusions**

In summary, we have shown that for shear horizontal surface acoustic wave scattering into bulk elastic waves, assuming self-affine or power-law roughness, the associated quality factor is strongly influenced by the corresponding roughness exponent H and the amplitude to correlation length ratio. In any case, the quality factors obtained for *H≥0.5* is comparable to quality factors due to intrinsic dissipation mechanisms due to thermoelastic and phonon-phonon scattering especially for wave vector ranges *kξ<1*, as a such it should be carefully considered in estimating dissipation loss in modern MEMS/NEMS. Further studies are in progress to understand further contributions arising from other modes of surface sound waves such as the Reyleigh surface waves that also undergo scattering by surface roughness, however, with distinct features than SH waves (Love waves) considered in this study [16].

**Acknowledgements**

I would like to acknowledge useful discussions with A. A. Maradudin.

**Figure captions**

**Figure 1 (a)** Schematic showing the propagation geometry, and **(b)** showing the influence of the roughness exponent H for three surfaces with the same *w* and *ξ*.

**Figure 2** Calculations of $\omega_{1,2}$ vs. *x* (=$k\xi$) for two different roughness exponents, *w*=3 *nm*, *ξ=60 nm*, and $a_o$=0.3 *nm*.

**Figure 3** Calculations of $Q_s$ vs. *x* (=$k\xi$) for various roughness exponents *H*, w=3 nm, *ξ=60 nm*, and $a_o$=0.3 *nm*.

**Figure 4** Calculations of $Q_s$ vs. x (=$k\xi$) for various lateral roughness correlation lengths, *H=0.7, w=3 nm*, and $a_o$=0.3 *nm*. The inset shows calculations of $Q_s$ vs. *x* (=$k\xi$) for various roughness amplitudes *w, ξ=60 nm, H=0.7*, and $a_o$=0.3 *nm*.

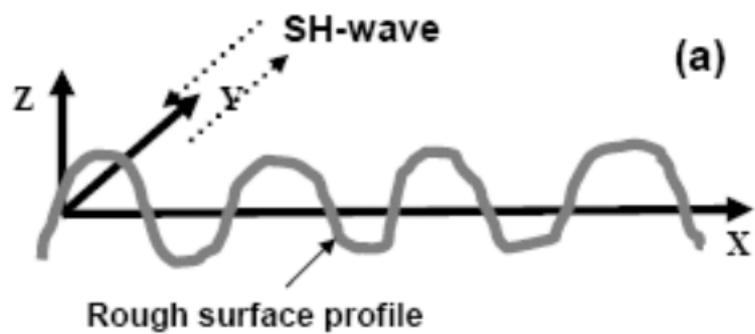

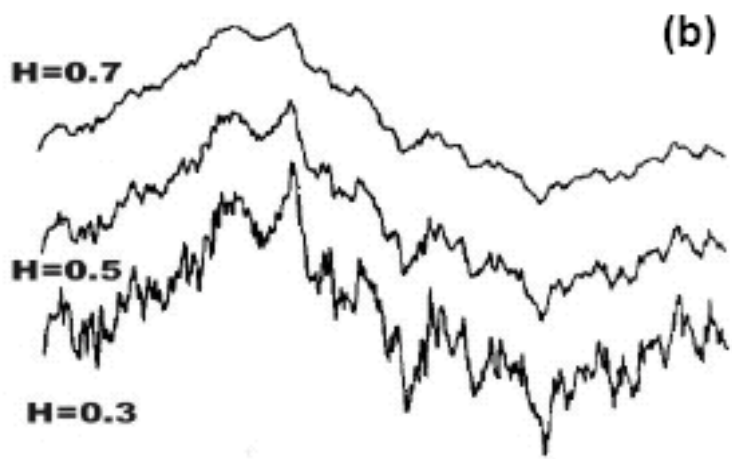

**FIGURE 1**

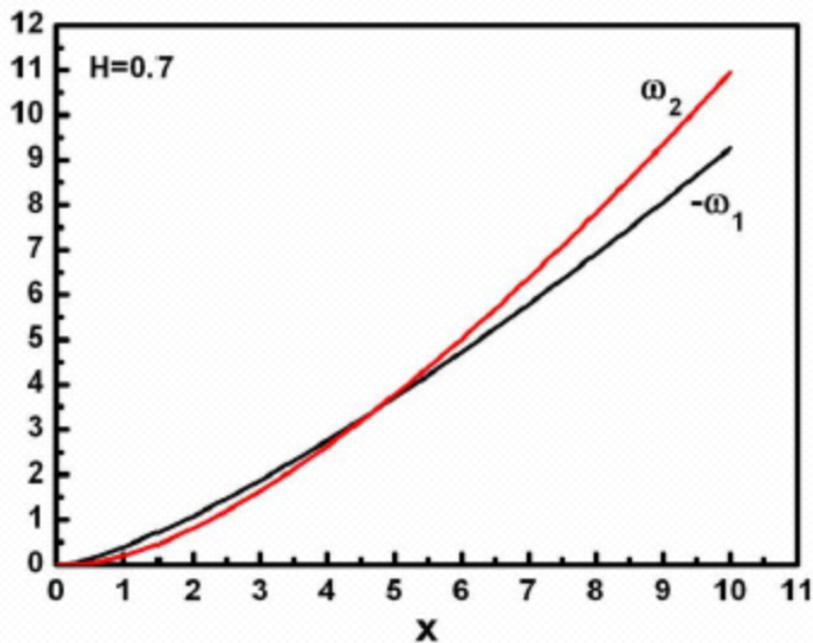
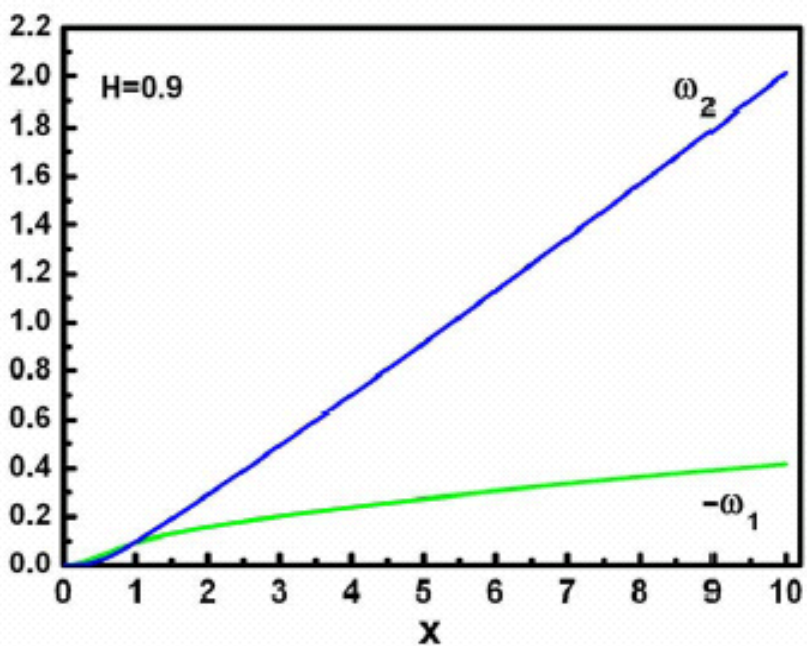

FIGURE 2

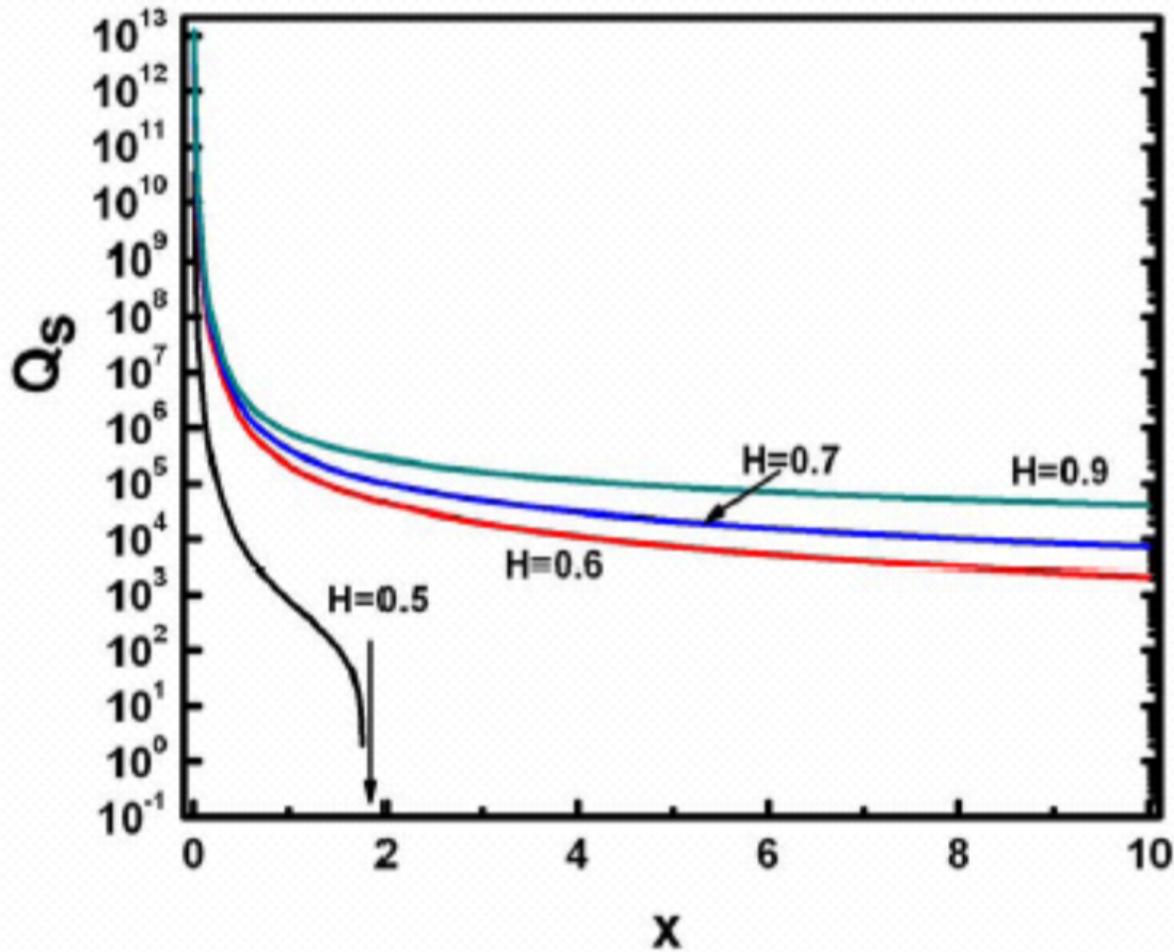

FIGURE 3

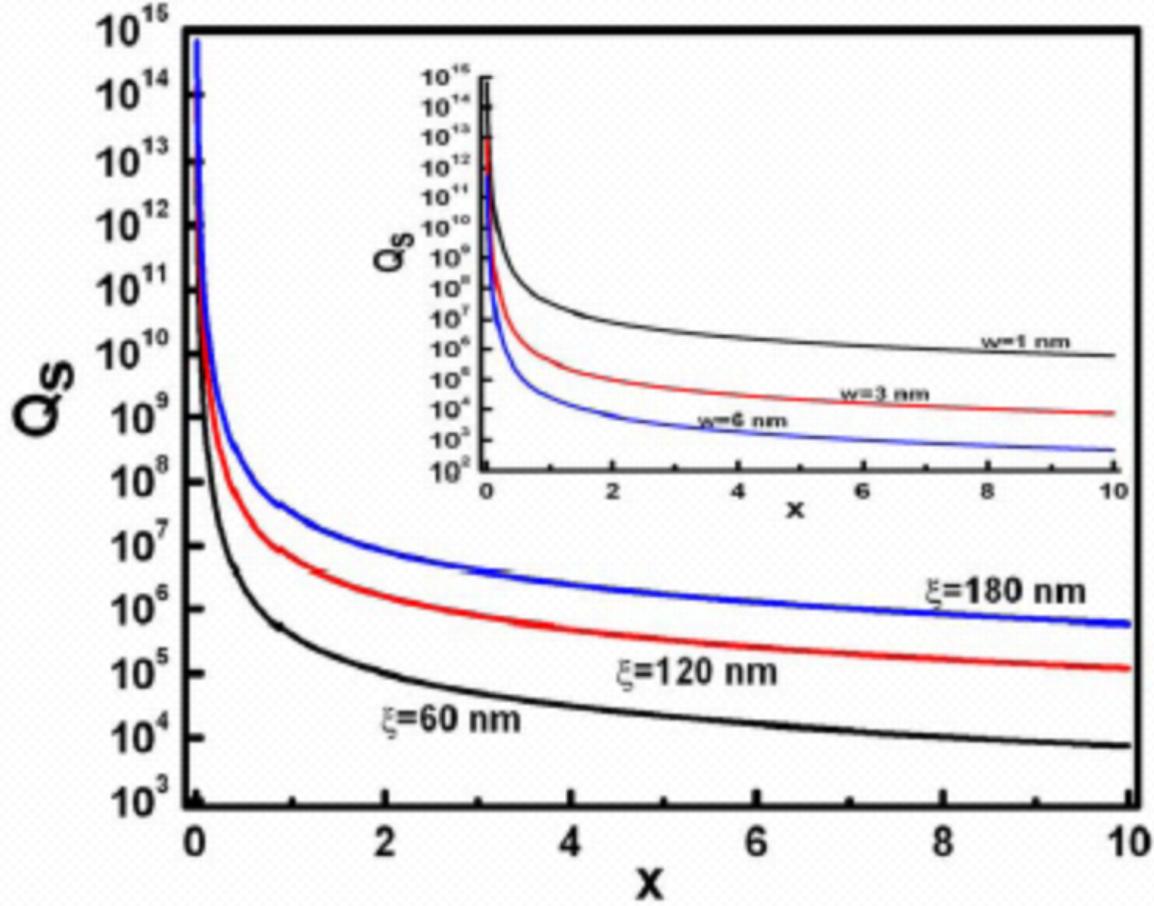

FIGURE 4